\documentclass[10pt,twocolumn]{article}

\usepackage[utf8]{inputenc}
\usepackage[T1]{fontenc}
\usepackage{microtype}
\usepackage{authblk}

\usepackage{times}
\usepackage{datetime}
\usepackage{graphics} 
\usepackage{url}
\usepackage{hyperref}
\usepackage{color}
\usepackage{textcomp}
\usepackage{booktabs}
\usepackage{ccicons}
\usepackage[tight,footnotesize]{subfigure}
\usepackage{pifont}
\usepackage{multirow}
\usepackage[justification=centering]{caption}
\usepackage{xspace}
\usepackage{comment}
\usepackage{tikz}
\usepackage{flushend}
 
\usepackage{xcolor}

\newcommand{\IdealReplay}{\texttt{iRe\-play\-er}}

\newcommand{\IR}{\texttt{iRe\-play\-er}}

\newcommand{\specialcell}[2][c]{%
  \begin{tabular}[#1]{@{}c@{}}#2\end{tabular}}

\definecolor{Gray}{cmyk}{0,0,0,0.5}

\date{}

\begin{document}

\title{\IdealReplay{}: In-situ and Identical \\Record-and-Replay for Multithreaded Applications}

\author[1]{Hongyu Liu}
\author[1]{Sam Silvestro}
\author[1]{Wei Wang}
\author[2]{Chen Tian}
\author[1]{Tongping Liu}
\affil[1]{University of Texas at San Antonio, USA}
\affil[2]{Huawei US R\&D, USA}
\affil[ ]{\textit{liuhyscc@gmail.com, Sam.Silvestro@utsa.edu,} \newline \textit{Wei.Wang@utsa.edu, Chen.Tian@huawei.com, Tongping.Liu@utsa.edu}}

\maketitle

\begin{abstract}
Reproducing executions of multithreaded programs is very challenging due to many intrinsic and external non-deter\-ministic factors. Existing RnR systems achieve significant progress in terms of performance overhead, but none targets the in-situ setting, in which replay occurs within the same process as the recording process. Also, most existing work cannot achieve identical replay, which may prevent the reproduction of some errors. 

This paper presents \IR{}, which aims to \textbf{i}dentical\-ly \textbf{replay} multithreaded programs in the original process (under the ``in-situ'' setting). The novel in-situ and identical replay of \IR{} makes it more likely to reproduce errors, and allows it to directly employ debugging mechanisms (e.g. watchpoints) to aid failure diagnosis. Currently, \IR{} only incurs 3\% performance overhead on average, which allows it to be always enabled in the production environment. \IR{} enables a range of possibilities, and this paper presents three examples: two automatic tools for detecting buffer overflows and use-after-free bugs, and one interactive debugging tool that is integrated with \texttt{GDB}. 

\end{abstract}

{\bf Keywords:} Record-and-Replay, Identical Replay, In-situ Replay, Multithreaded Debugging

\section{Introduction}
\label{sec:intro}

Multithreaded programs contain intrinsic non-deterministic factors that may affect the schedule and results of different executions. Thus, reproducing multithreaded programs is very challenging. Record-and-Replay (RnR) systems record non-deterministic events of the original execution, such as the order of synchronizations and the results of certain system calls, and then reproduce these events during the re-execution~\cite{PRES}. Some RnR systems even record the order of memory accesses~\cite{Bhansali:2006:FIT:1134760.1220164}, or utilize offline analysis to infer the order of memory accesses inside the execution~\cite{ODR, Huang:2013:CRL:2491956.2462167, Liu:2015:LRV:2737924.2738001, huang2017towards}. However, existing RnR systems have two shared shortcomings, in addition to their specific problems as described in Section~\ref{sec:relatedwork}. 

First, they do not support in-situ replay, typically reproducing the execution in a different process. They could possibly achieve better diagnostic capability, since they can access all information from the entire execution~\cite{Arulraj:2014:LSM:2541940.2541973}. However, there are several issues. (1)~As observed by experts~\cite{Tucek:2007:TDP:1294261.1294275, Kasikci:2015:FST:2815400.2815412}, offline replay requires the same runtime environment as the recording process, which will greatly limit their usage, since normal users may not want to share third-party libraries or sensitive inputs/logs with programmers due to business and privacy concerns. (2)~They cannot be utilized to assist online recovery~\cite{Tucek:2007:TDP:1294261.1294275}. 

\begin{figure}[!h]
\begin{center}
\includegraphics{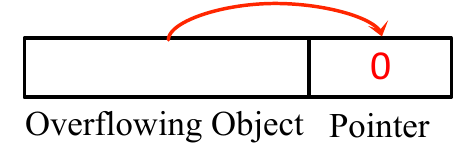}
\end{center}
\vspace{-0.1in}
\caption{
A null reference problem in original execution. \label{fig:nullreference}}
\vspace{-0.1in}
\end{figure}

Second, most existing RnR systems (except RR~\cite{RR,RRPaper}) cannot identically reproduce the recorded execution, as they do not guarantee the same system states, such as process/{\allowbreak}thread IDs and file descriptors~\cite{Ronsse:1999:RFI:312203.312214, PRES, ODR, DoublePlay, Hower:2008:REE:1381306.1382144, Montesinos:2008:DRD:1381306.1382146, Honarmand:2013:CUA:2451116.2451138, Honarmand:2014:RDL:2665671.2665737, Huang:2013:CRL:2491956.2462167, Bond:2015:EDR:2807426.2807434, Liu:2015:LRV:2737924.2738001, Castor}, the same results of system calls (e.g., \texttt{time})~\cite{huang2010leap, Dthreads, Huang:2013:CRL:2491956.2462167}, or have different memory layouts~\cite{Ronsse:1999:RFI:312203.312214, PRES, ODR, DoublePlay, Hower:2008:REE:1381306.1382144, Montesinos:2008:DRD:1381306.1382146, Honarmand:2013:CUA:2451116.2451138, Honarmand:2014:RDL:2665671.2665737, Huang:2013:CRL:2491956.2462167, Bond:2015:EDR:2807426.2807434, Liu:2015:LRV:2737924.2738001, Castor}. 
Therefore, it is impossible to reproduce some types of bugs: (1)~Bugs related to memory layout may not be reliably reproduced. Figure~\ref{fig:nullreference} shows such an example, in which a crash occurs when dereferencing a null pointer caused by a buffer overflow. A different memory layout, where an integer (not the pointer) is allocated immediately after the overflowing object, may hide this crash during re-executions; (2)~Bugs dependent on system states, such as thread IDs, file descriptors, or memory addresses, may not be reproducible, when the states in replay are not the same as those in the original execution. However, many real systems were designed to utilize system states explicitly. For instance, the Hoard allocator assigns heaps to each thread based on the hashing of their thread ID~\cite{Hoard}, and some hash tables use object addresses as their keys~\cite{hashkey}. 

This paper presents \IR{}, a novel system that targets to in-situ and identically replay multithreaded programs, which has the following significant differences from existing RnR systems. 

\textbf{First, iReplayer designs an in-situ replay technique that always replays the last-epoch execution within the same process as the original execution}. The in-situ replay makes it easier to replay identical system states and is more likely to reproduce bugs.
This in-situ replay is different from existing online replay~\cite{Rx, Tucek:2007:TDP:1294261.1294275, Lee:2010:REO:1736020.1736031, DoublePlay}, where their replays actually occur in a process different from the recorded one. Currently, \IR{} only replays the last-epoch execution by default. However, it is especially suitable for identifying bugs. Based on recent studies~\cite{1209956, Rx, Arulraj:2014:LSM:2541940.2541973}, most bugs have a very short distance of error propagation, which indicates a root cause may be located shortly prior to failures. Replaying-last-epoch also avoids significant time spent waiting for problems to appear.

\textbf{Second, iReplayer aims for identical re-execution that strictly preserves all system states, results of system calls, the order and results of synchronizations, and the same memory allocations/deallocations of the original execution, even for racy applications}. Re-execution in the same process as the original execution helps preserve system states, such as process IDs. Additionally, \IR{} handles system calls specially, delays the reclaiming of threads in order to maintain the state of memory mappings and IDs for each thread, and employs a custom memory allocator to manage the application heap similarly across multiple executions, as described in Section~\ref{sec:identicalchallenges}. Based on our evaluation, \IR{} can identically reproduce all evaluated applications (even racy ones) that do not contain implicit synchronizations (i. e., without using pthread APIs).

\textbf{Thirdly, iReplayer only imposes 3\% recording overhead on average, which is sufficiently low for deployment}. \IR{} utilizes multiple approaches to reduce its logging overhead: (1)~it takes advantage of the in-situ setting to avoid recording the content of file reads/writes; (2)~It avoids the recording of memory accesses by handling race conditions in replay phases, inspired by existing work~\cite{Lee:2014:ILR:2945642.2945655,Castor}; (3)~It avoids the recording of memory allocations by employing a novel heap design, inspired by Dthreads~\cite{Dthreads}; (4)~It also designs a novel data structure that allows it to efficiently record the local-order of synchronizations, while still ensuring identical replay; (5)~\IR{} designs an indirect level for recording synchronization events, similar to existing work~\cite{SyncPerf}. These are the major reasons why \IR{} has much less overhead than past related techniques--e.g. Respec~\cite{Lee:2010:REO:1736020.1736031}. More details can be seen in Section~\ref{sec:normalexecution}. 

\textbf{The identical and in-situ re-execution of iReplayer enables a range of possibilities, and three tools are shown in this paper}. These tools can be utilized in staging or canary deployment, especially when new features are rolling out. In addition, the in-situ and identical replay of \IR{} enables unique possibilities: (1)~It enables on-site tools that can automatically diagnose root causes of program failures, such as memory errors, segmentation faults, aborts, and assertions. For instance, upon faults, we could perform binary analysis to pinpoint faulting addresses, then install watchpoints on them to identify root causes on-site without human involvement. In contrast, offline RnR cannot perform on-site analysis, and typically require additional human effort. (2)~It enables evidence-based approaches to prevent program failures, such as memory errors or deadlocks. For instance, it is possible to extend \IR{} to delay memory deallocations to prevent discovered use-after-frees, or enforce an alternative lock order to avoid deadlocks. It is impossible to perform online repair with existing offline RnRs.

\noindent Overall, this paper makes the following contributions:

\paragraph{First in-situ record-and-replay technique for multithreaded programs:} \IR{} proposes the first in-situ RnR system that the replay occurs in the same process as the original execution, enabling new possibilities. 

\paragraph{An identical replay technique:} \IR{} supports the identical replay of multithreaded programs without self-de\-fined synchronizations. The identical replay helps reproduce bugs, and ease the development of automatic tools. 

\paragraph{Practical implementation techniques to reduce overhead:} \IR{} makes multiple design choices to reduce recording overhead: it proposes a novel data structure that supports identical replay with low recording overhead, and supports the checking of divergence easily during the replay; it designs a novel heap to avoid the recording of memory allocations.  

\paragraph{A practical system combining low recording overhead and convenience:} 
\IR{} is a software-only solution with negligible recording overhead, only 3\% on average. \IR{} is a drop-in library that runs entirely within the user space, and does not require nonexistent hardware, customized OS, or the modification of programs.  

\paragraph{Multiple promising applications:} To demonstrate the usefulness of \IR{}, this paper developed two tools to detect heap over-writes and use-after-free errors, and one interactive debugging tool (connecting with \texttt{GDB}). 

\subsection*{Outline:}
The remainder of this paper is organized as follows. Section~\ref{sec:overview} gives an overview of our approach, including the challenges of implementing multithreading support. After that, Section~\ref{sec:implementation} presents the detailed implementation, and Section~\ref{sec:applications} discusses several applications built on \IR{}.
Section~\ref{sec:evaluation} presents experimental results, and limitations are discussed in Section~\ref{sec:discussion}. Finally, Section~\ref{sec:relatedwork} reviews related work, and Section~\ref{sec:conclusion} concludes.

\section{Overview}

\label{sec:overview}

\begin{figure*}[!t]
\begin{center}
\includegraphics[width=4.5in]{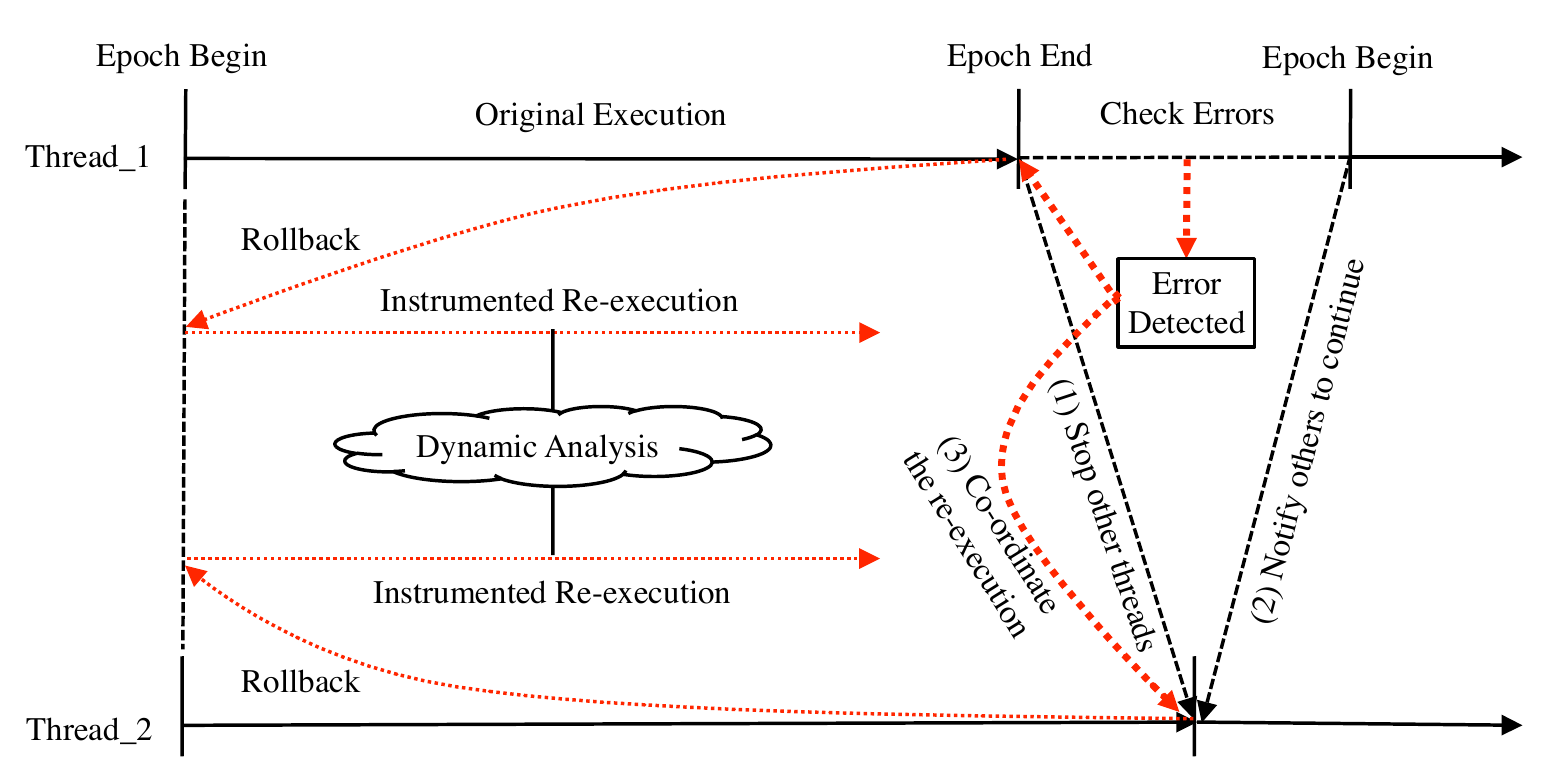}
\end{center}
\vspace{-0.1in}
\caption{
Overview of \IR{} with two threads. Bold lines represent normal program executions, and dashed lines illustrate the assistance functions at epoch boundaries. Dash-dot lines and red short-dash lines are tool-specific functions. 
\label{fig:multioverview}}
\vspace{-0.1in}
\end{figure*}

This section provides an overview of \IR{}, and the major challenges of supporting in-situ and identical replaying multithreaded applications. 

\subsection{Overview of Execution}
\IR{} divides the entire execution into multiple ep\-ochs, based on irrevocable system calls (defined in Section~\ref{sec:identicalchallenges}), abnormal exits, or user-defined criteria. For instance, users may use the size of logging as the criteria, in order to reduce memory/disk consumption. 

The overview of \IR{} is illustrated in Figure~\ref{fig:multioverview}, which shows an execution with two threads. At the beginning of an epoch, \IR{} takes a snapshot of the program's states, such as its memory and the position of open files, so that the program can be rolled back to this point (Section~\ref{sec:epochbegin}). During the original execution, \IR{} records the order of synchronizations (Section~\ref{sec:normalexecution}), and handles system calls differently (Section~\ref{sec:syscalls}). When a thread encounters an irrevocable system call -- which changes the state, but cannot be safely rolled back -- or reaches the user-defined criteria for recording, it will be treated as the coordinator thread, and will coordinate with other threads to pause the execution (Section~\ref{sec:epochend}). After all threads have reached a quiescent state, the coordinator thread determines whether to continue the execution, or perform re-execution, based on user instructions or tool-specific evidence (see Section~\ref{sec:applications}). If a replay is required, the coordinator notifies all other threads to roll back (Section~\ref{sec:rollback}) and re-execute the program (Section~\ref{sec:reexecutions}). Otherwise, all other threads are notified to proceed to the next epoch.

\subsection{Challenges for In-situ and Identical Replay} 

\label{sec:identicalchallenges}

The major challenge of \IR{} is to ensure identical replay of synchronizations, memory accesses, system calls, and memory layout under the in-situ setting. 

\subsubsection{Ensuring Identical Synchronizations} 

We record the order of synchronizations during the original execution, then replay this order in re-executions~\cite{PRES}. However, the greatest challenge is to achieve efficient recording, which is further described in Section~\ref{sec:normalexecution}. Since some applications rely on the results of synchronization functions, such as try locks or barrier waits, \IR{} also records the return values of synchronizations and returns them in replays. 

Currently, \IR{} does not support programs with ad hoc synchronizations, where programs use their own synchronization methods rather than explicit pthread APIs~\cite{Xiong:2010:AHS:1924943.1924955}), as further discussed in Section~\ref{sec:discussion}.

\subsubsection{Ensuring Identical Order of Memory Accesses} 
\label{sec:race}
Two types of memory accesses exist in multithreaded applications, including thread-private and shared accesses. The order of thread-private accesses is determined by the order of instructions, which does not require special handling for identical replay. Shared accesses will be identical if they are properly protected by explicit synchronizations, when explicit synchronizations are identically reproduced. Thus, the difficulty lies in ensuring the identical replay for race conditions.

\paragraph{Handling race conditions:} \IR{} does not record rac\-y accesses initially, since that is too expensive~\cite{Bhansali:2006:FIT:1134760.1220164}. Instead, it handles race conditions inside replay phases, which avoids significant recording overhead for common cases in which programs do not expose race conditions. During replay, it will check for the divergence from the recorded events. If a replay behaves exactly the same as the original execution, i.e. the same order of system calls and synchronizations, then race conditions are either not exposed or are successfully reproduced. Otherwise, \IR{} immediately initiates another replay, and utilizes multiple replays to search for a matched schedule. When the events of a replay match the recorded ones, \IR{} assumes that the replay is identical to the original execution, and will stop searching. To our understanding, the chance that the replay is still different from the original, but with the same order of schedules, is very low. This general idea is inspired by existing work~\cite{Lee:2014:ILR:2945642.2945655, Castor}, but with some difference. Lee et al. utilize a single-threaded execution to replay multithreaded applications~\cite{Lee:2014:ILR:2945642.2945655}.

\subsubsection{Ensuring Identical System Calls} 
\label{sec:syscalls}
Existing RnR systems record the results of system calls, and replay the same states during replay~\cite{RecPlay, PRES}. For instance, the recorded results of \texttt{gettimeofday} will be returned in the replay phase. However, no existing work aims for the in-situ setting.
The in-situ setting imposes some additional challenges toward ensuring identical system calls. For instance, we assume there is a sequence of file-related system calls, such as \{\texttt{open(1)}, \texttt{close(1)}, \texttt{open(2)}\}, where \texttt{open(2)} will return the same file descriptor as \texttt{open(1)}. If this sequence of system calls is later replayed, it is impossible to ensure the same file descriptor for \texttt{open(1)}, since \texttt{open(2)} now occupies it. Similar results may occur for the \texttt{munmap} system call. \IR{} classifies system calls into five categories, similar to DoubleTake~\cite{DoubleTake}. 

\textbf{Repeatable system calls} always return the same results within the in-situ setting, e.g. \texttt{getpid()}. They require no special handling in either the recording or replaying phases. 

\textbf{Recordable system calls} return different results when invoked during re-executions, such as \texttt{gettimeofday()} and socket reads/writes. \IR{} records the results, and returns the same values during replay without actual invocations. 

\textbf{Revocable system calls} modify system states, but the results of these operations can be reproduced identically under the in-situ setting, as long as initial states are recovered before the re-execution. These system calls mainly include file-related reads/writes. Although their results can be recorded, this may impose substantial recording overhead~\cite{Lee:2010:REO:1736020.1736031}. Instead, \IR{} records the positions of open files during the recording phase, and issues these system calls normally during replays (after recovering positions). However, if a \texttt{write} changes the data after invoking \texttt{lseek}, then it is unable to reproduce any \texttt{read} prior to the \texttt{lseek}. Therefore, \IR{} treats \texttt{lseek} (with repositions) as an irrevocable system call. 

\textbf{Deferrable system calls} irrevocably change system states, but can be safely delayed. These system calls, such as \texttt{mun\-map} and \texttt{close}, are very important for identical re-execution in an in-situ setting. \IR{} delays these system calls until the next epoch, when there is no need for re-execution. Note that delaying \texttt{close()} may result in the number of open files exceeding the default limit; therefore, \IR{} increases this limit during initialization. 

\textbf{Irrevocable system calls} irrevocably change system states, and cannot be rolled back safely or deferred easily. Although conceptually they can be recorded as in other existing RnR systems~\cite{PRES, ODR, Castor}, this may involve substantial performance overhead or engineering effort. Currently, \IR{} simply treats them
as irrevocable system calls, and closes the current epoch when encountering them. For instance, \texttt{execve} and \texttt{fork} are examples of such system calls. 

This classification has a significant impact on performance. Although \IR{} could treat every system call as irrevocable, this would create a large number of epochs, and significantly increase the overhead caused by stopping, checkpointing, and cleaning upon epochs. Thus, irrevocable systems calls are eliminated as much as possible. 
Some system calls are further classified based on their input parameters. For instance, the \texttt{fcntl} system call with the \texttt{F\_GETOWN} flag is treated as a repeatable system call, while it will be treated as a recordable system call when used with the \texttt{F\_DUPFD} flag. 

\subsubsection{Ensuring Identical Heap Layout}
\label{sec:mtheap}
Memory management is a major source of non-determinism in multithreaded applications. First, the OS may randomize memory uses due to the ASLR mechanism~\cite{ASLR}. Second, multiple threads may compete with each other. To ensure the identical memory layout, \IR{} isolates its internal memory uses from those of applications, adapts a ``per-thread heap'' so that memory allocations inside the same heap completely depend on the program order, and controls interactions among different threads. 

\textbf{Per-thread Heap:} Built on top of HeapLayers~\cite{heaplayers}, it adapts the per-thread heap organization of Hoard~\cite{Hoard}. Memory allocations and deallocations within each thread will be identically reproduced, if they do not interfere with other threads. Different from Hoard, two live threads are never allocated from the same per-thread heap.
\IR{} intercepts thread creation, and deterministically assigns a unique heap for every thread by utilizing a global lock to serialize thread creation. When the order of locks is replayed deterministically, each thread will have the same heap during re-executions. 

\textbf{Deterministically fetches blocks for per-thread heaps:} \newline \IR{} maintains a super heap that holds a large number of blocks for all per-thread heaps. When a per-thread heap exhausts its memory, it obtains a new block from the super heap under the protection of a global lock, which is guaranteed to be the same during re-executions via deterministically replaying lock acquisitions.  

\textbf{Handles deallocations by a different thread deterministically:} \IR{} always returns a freed object to the current thread issuing the \texttt{free}, no matter which thread allocated the object initially. Thus, this freed object only affects the subsequent memory allocations of the current thread, which again depends on the program order, and will be deterministic. 

Inside each per-thread heap, objects are managed using power-of-two size classes. During allocations, each request will be aligned to the next power-of-two size. The free list will be checked first, and only if the request cannot be allocated from its free list, it will be allocated using the bump pointer mechanism~\cite{Berger:2002:RCM:582419.582421}. 
Upon deallocation, each deallocated object will be inserted into the head of its corresponding free list and will be reutilized consequently. 
Since \IR{} limits memory allocations to its per-thread heap and controls the interactions among different threads, there is no need to record the addresses of allocations to ensure identical replay. Note that \IR{} does not serialize memory allocations, but only the acquisition of each block (4 megabytes). Instead, \IR{} avoids the usage of locks upon each allocation, which explains why its heap is 3\% faster than the default Linux allocator (Section~\ref{sec:performance}).

\subsection{Other Challenges}
\label{sec:otherchallenges}

There are other challenges, mostly caused by the in-situ setting: how to perform recording efficiently (Section~\ref{sec:normalexecution})? How to stop an epoch safely under the in-situ setting (Section~\ref{sec:epochend}), when some threads may be in the middle of a system call or waiting for synchronizations? How to roll back multiple threads correctly, especially for threads created in the last epoch or are waiting on synchronizations (Section~\ref{sec:rollback})? How to prepare for re-execution (Section~\ref{sec:rollback}) to assist identical replay? How to control the order of re-executions, and detect divergence possibly caused by race conditions (Section~\ref{sec:reexecutions})? 

\section{Implementation}
\label{sec:implementation}

\begin{figure*}
    \centering
    \begin{minipage}{0.4\textwidth}
        \centering
        \includegraphics[width=0.85\textwidth]{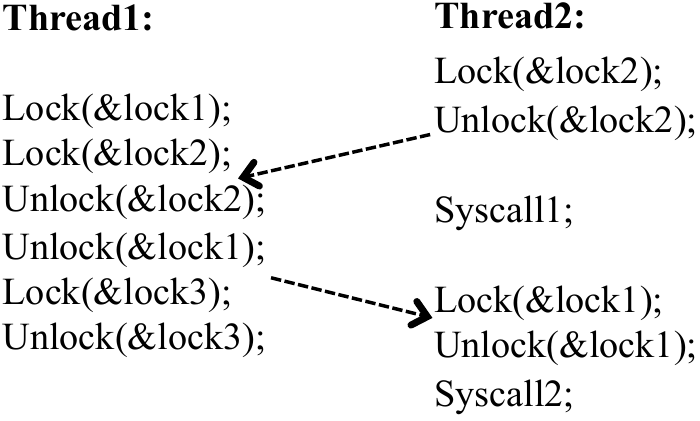} 
        \vspace{-0.1in}
        \caption{Code snippet with sync and syscalls \label{fig:code}}
        \vspace{-0.1in}
    \end{minipage}\hfill
    \begin{minipage}{0.45\textwidth}
        \centering
        \includegraphics[width=0.9\textwidth]{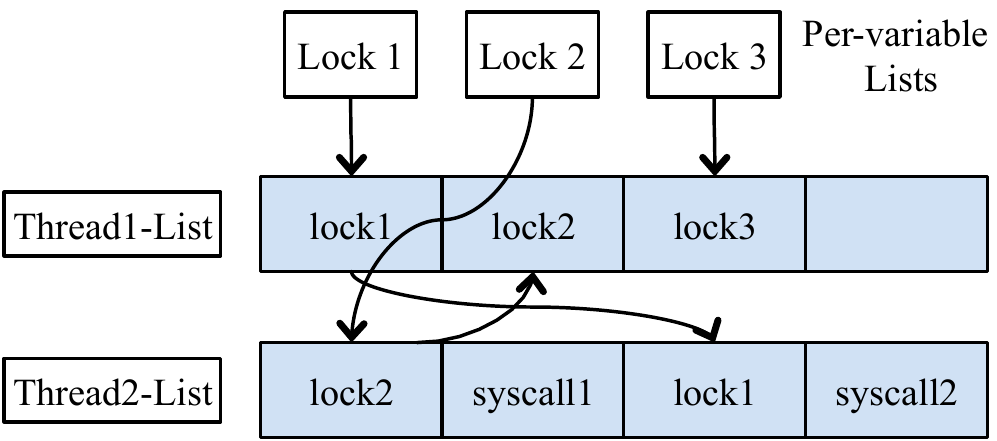} 
        \caption{Data structures for tracking events \label{fig:structures}}
        \vspace{-0.1in}
    \end{minipage}
\end{figure*}

This section describes the implementation of \IR{}, organized by phases that are shown as Figure~\ref{fig:multioverview}.

The start of a program is considered the start of the first epoch, and terminations (either normal or abnormal exits) will be treated as the end of the last epoch. \IR{} marks its initialization function with the \texttt{constructor} attribute, which allows it to initialize its custom heap, install signal handlers, and prepare internal data structures for recording, before entering the \texttt{main} routine. During initialization, \IR{} identifies the range of global and text segments for the application, as well as any libraries, by analyzing the \texttt{/proc/self/maps} file. This information will be utilized for checkpointing in the original execution, or for preparation for re-executions.

\subsection{Epoch Begin}
\label{sec:epochbegin}

At epoch begin, the major task is to checkpoint the states of the execution in order to support re-executions. If the epoch is not the first one, some housekeeping operations should be completed prior to checkpointing. In a multithreaded environment, a thread (typically the coordinator thread) is responsible for the housekeeping operations. 

Housekeeping operations typically involve the removal of unnecessary records from the previous epoch, such as the list of system calls and synchronizations. As described in Section~\ref{sec:syscalls}, some system calls are delayed, such as \texttt{close} and \texttt{munmap}, which will be issued at this time. Cached data for closed sockets will be removed, and joined threads will be reclaimed. 

After this, \IR{} checkpoints the states shared by all threads, such as the memory states, and positions of open files (see Section~\ref{sec:normalexecution}). Checkpointing memory states is performed by copying all writable memory to a separate block of memory, such as the heap and globals for both the application and its dynamically-linked libraries. This checkpointing does not require changes to the underlying operating system, which is different from existing work~\cite{Rx, Tucek:2007:TDP:1294261.1294275, Lee:2010:REO:1736020.1736031, DoublePlay}. \IR{} also updates file positions of all open files, which is tracked in a global hash table.

Afterwards, all other threads are woken up, including th\-reads waiting on condition variables, barriers, and thread joins, so that they can checkpoint their own per-thread states. Per-thread states include the stack and per-thread hardware registers. \IR{} invokes \texttt{getcontext} to record the state of per-thread hardware registers. 

\subsection{Original Execution}

\label{sec:normalexecution}
During the original execution, \IR{} mainly handles system calls and synchronizations, and deals with memory allocations and deallocations as discussed in Section~\ref{sec:mtheap}. \IR{} utilizes the following mechanisms to reduce recording overhead.

First, \IR{} designs a novel data structure (as shown in Figure~\ref{fig:structures}) to store synchronization and system call events, which preserves the order of events in the same thread and across multiple threads. Each event is recorded in its per-thread list initially, then will be added into the corresponding per-variable list. For the example shown in Figure~\ref{fig:code}, the corresponding order will be recorded as in Figure~\ref{fig:structures}. For instance, $lock1$'s per-variable list will track that $lock1$ is first acquired by Thread1, and then by Thread2.  

This data structure removes the need for the global order~\cite{PRES}, reconstructing the synchronization order offline~\cite{Huang:2013:CRL:2491956.2462167, huang2017towards, Castor}, and special hardware support~\cite{Castor}. It guarantees identical re-execution of explicit synchronizations: different synchronizations inside the same thread will always have the same order, determined by its program logic; the per-variable list ensures that multiple threads will perform synchronizations in the recorded order. In addition, this data structure is convenient for checking the divergence during re-executions: each thread is only required to check whether the next event is the same as the recorded one in their per-thread lists. If not, this is an indication of a divergence, and \IR{} will immediately invoke a re-execution in order to search for a matching schedule. 

This data structure is very efficient at tracking events: (1)~Its recording does not introduce new lock contention between threads, excepting the original lock operations, which is different from existing work~\cite{Lee:2010:REO:1736020.1736031}. (2)~\IR{} further reduces its logging overhead by pre-allocating a specified number of entries for per-thread lists, which does not require additional memory allocations during the recording. When all entries are exhausted, it is time to stop the current epoch and start a new epoch.  

Second, \IR{} employs a level of indirection for finding the list associated with each synchronization variable, instead of naively using a global hash table, similar to SyncPerf~\cite{SyncPerf}. It is difficult to define the size of the hash table and design a balanced hash algorithm. The naive method was found to impose up to $4\times$ performance overhead when applications has a large number of synchronization variables, e.g. \texttt{fluidanimate} of PARSEC~\cite{parsec}. Instead, \IR{} allocates a shadow synchronization object from its internal heap (to avoid interfering with the application's memory uses), and saves the pointer to this shadow object within the first word of the original synchronization object. This shadow object includes the real synchronization object and a pointer to its per-variable list. The in-direction level allows for quickly obtaining per-variable lists within a few operations. 

Third, \IR{} avoids the recording of each memory access by delaying the handling of race conditions until the replay phase (Section~\ref{sec:race}), avoids the recording of memory allocations by employing a novel heap (Section~\ref{sec:mtheap}), and avoids the recording of file reads/writes by treating them as revocable system calls (Section~\ref{sec:syscalls}).

\subsubsection{Supporting Synchronizations}
\IR{} supports a range of synchronization primitives, such as thread creations, various forms of mutex locks, condition variables, barriers, signals, and thread joins. 

\paragraph{Thread creation, destruction, and joins:} \IR{} intercepts \texttt{pthread\_create} function calls to initialize thread-related data, checkpoint the state prior to the execution, and handle future thread exits. It takes multiple approaches to guarantee identical replay: (1)~It does not allow concurrent thread creations by using a global mutex; (2)~\IR{} keeps threads alive (without exiting) until the next epoch in order to preserve system states, such as thread IDs and stacks, by using a thread-specific conditional variable and a status field. For a joinee thread joined by its parent, it checks this status field upon exit. If the parent thread has not yet joined on it, the status will be set to ``joinable'', which may then be changed to ``joined'' with a subsequent join operation. Otherwise, the joinee thread wakes its joiner immediately. Joinee threads are always waiting on a condition variable (and thus kept alive), awaiting notification to either roll back or exit. 

\paragraph{Mutex locks:} For mutex locks, \IR{} records the order and return values of lock acquisitions using the data structures shown in Figure~\ref{fig:structures}. For mutex try-locks, \IR{} also records the return value within per-thread lists, but only adds successful acquisitions into per-variable lists.

\paragraph{Condition variables:} A \texttt{cond\_wait} is treated as a mutex release followed by a mutex acquisition (when woken up). \IR{} records the wake-up events of condition variables, similar to lock acquisitions. Since other threads may close the current epoch during waiting, every thread should record its state and which condition variable before waiting. After being woken up, a thread either proceeds as normal, or performs a re-execution or checkpointing. \IR{} does not record the order of \texttt{cond\_signal} and \texttt{cond\_broadcast}, but only the wake-up order of threads. Note that this method may induce a non-identical replay when locks are not properly acquired. \IR{} overcomes this by inserting random sleeps at diverging points, as shown in Section~\ref{sec:identicalevaluation}. 

\paragraph{Barriers:} 
During barrier waiting, a thread may wait inside the kernel until the required number of threads have reached the same barrier. However, it is difficult to wake up a thread waiting on the barrier. To solve this issue, \IR{} re-implements the barrier by combining a mutex and a condition variable, as it is easier to wake up a thread waiting on a condition variable using a signal. \IR{} intercepts the initialization of barriers in order to initialize the corresponding mutex locks and condition variables. \IR{} does not record the order of entries into a barrier, since a thread waiting on a barrier will not change the state. Instead, \IR{} records the return value of every barrier wait, since some applications may rely on it.  

\subsection{Epoch End}
\label{sec:epochend}
At epoch end, the coordinator thread is responsible for safely stopping the other threads and closing the current epoch.
It is impossible to checkpoint a multithreaded program correctly and replay identical consequently, when multiple threads continue executing and changing states. Therefore, \IR{} adapts the ``stop the world'' approach employed by garbage collection~\cite{Boehm:1991:MPG:113445.113459}. Stopping an epoch safely is unique to the in-situ setting, which has several challenges as described below.

\textbf{Challenge 1:} How to stop other threads safely? Asynchronous methods (e.g. signals) are unreliable to stop and roll back threads cleanly, if these threads are waiting inside the kernel due to synchronizations (such as \texttt{barrier\_wait}) or other system calls. Instead, \IR{} employs a synchronized method: before any synchronization or system call invocation, it checks whether a coordinator thread has requested to stop the current epoch. If so, the current thread will wait on its internal condition variable, and mark its state as stopped. 

\textbf{Challenge 2:} How to stop threads waiting on synchronizations, such as lock acquisitions or condition variables, or blocking on external system calls? Threads waiting on condition variables are considered to be in unstable states, since other threads may wake them up at any time. For those threads, \IR{} continues checking their states until all other active threads have reached their stable stopped states. \IR{} also handles threads waiting on the acquisition of mutex locks: the actual holder of a mutex will release its lock temporarily before it stops, so that the waiter can acquire the lock and stop stably. \IR{} guarantees the lock will be returned to the original holder when the program proceeds as normal, without causing any atomicity violations. Some blocking IOs will be turned into non-blocking IOs upon interceptions, e.g. adding the timeout value for \texttt{epoll\_wait}.  

When all other threads (except the coordinator thread) have been stopped, the current epoch is closed. Thus, the coordinator thread will check whether a replay should be performed. If evidence of a program error exists (as shown in Section~\ref{sec:applications}), or instructions have been received from the user, all threads are rolled back to the last checkpoint and re-executed from there. Otherwise, all threads proceed to the next epoch as normal (discussed in Section~\ref{sec:epochbegin}). The coordinator thread orchestrates these operations.

\subsection{Preparing for Re-execution}
\label{sec:rollback}

\IR{} prepares the re-execution in the following steps. 

Firstly, the coordinator thread restores the memory of heap and global sections for both the application and all shared libraries, by copying the backup memory back its corresponding locations.

Secondly, \IR{} resets pointers of all per-thread lists and per-variable lists to their first recorded entry. Internally, \IR{} maintains a hash table to track the mapping between synchronization variables and their shadow objects to assist this. 

Thirdly, \IR{} recovers file positions of all opened files from the last epoch, by invoking the \texttt{lseek} API directly with the \texttt{SEEK\_SET} option on every file descriptor. 

In the end, the coordinator instructs other threads to roll back their own stacks and contexts themselves. (1)~Threads waiting on condition variables or barriers should first be woken up. However, threads created during the last epoch should wait for notifications from their parents, after their corresponding thread-creation events have transpired. (2)~Rolling back the stack should be performed very cautiously, since the stack to be recovered might overwrite live values on the current stack, which can cause a program to behave abnormally. \IR{} forces all threads to use temporary stacks before copying, then switch back to their original stack after the copy has completed. 
(3)~Because \IR{} only restores the used portion of the stack in the last checkpoint, the remainder of the stack should be zeroed out to guarantee identical replay. Some applications contain un-initialized reads that may access stack variables beyond the stack of the last checkpoint. 
 (4)~If the rollback was caused by a program fault, such as \texttt{SIGSEGV}, \IR{} cannot perform the rollback directly inside the signal handler, which is using the kernel stack. For these cases, \IR{} passes control to a custom function by setting the IP pointer, so that the rollback can be performed in this function after returning from the signal handler. (5)~In the end, each thread calls the \texttt{setcontext} API to restore its hardware registers, and begins re-execution immediately thereafter. 

\subsection{Re-executions}
\label{sec:reexecutions}

\IR{}'s re-execution has three goals. First, it should identically reproduce the original execution. Second, it should check for possible divergence caused by race conditions. Third, it should handle signals triggered by watchpoints for applications, as described in Section~\ref{sec:applications}. 

\subsubsection{Repeating Original Execution}

To achieve identical re-executions, \IR{} handles system calls correspondingly (see Section~\ref{sec:syscalls}), repeats memory allocations and deallocations as described in Section~\ref{sec:mtheap}, and repeats the recorded order of synchronizations. Note that \IR{}'s replay is different from Castor~\cite{Castor}. Castor requires the construction of the order of synchronizations by using timestamps of different synchronizations. \IR{} designs a novel data structure (Section~\ref{sec:normalexecution}) to overcome this issue, which makes it suitable for in-situ setting.  

For each thread, \IR{} utilizes a condition variable and a mutex lock to control the re-execution, and relies on per-thread lists and per-variable lists to guide the re-execution (as described in Section~\ref{sec:normalexecution}). \textbf{The basic rule is listed as follows}: whenever the first event of a per-variable list (e.g., a lock) is also the first event of its corresponding per-thread list, the current thread can proceed. Otherwise, the current thread should wait on its condition variable until previous synchronizations on this variable have transpired.

\IR{} utilizes a global mutex to control the order of thread creations. During re-executions, the parent thread waits for its turn to proceed, then notifies the corresponding child thread (waiting on their internal condition variable) to proceed immediately. It skips the actual thread creation, since all threads were kept alive. This fact guarantees the same thread ID and stack for each thread. Other synchronizations are discussed in Section~\ref{sec:normalexecution}. 

\subsubsection{Checking Divergence}

\IR{} checks for the divergence from the recorded order for two types of events, system calls and synchronizations, using the data structure described in Section~\ref{sec:normalexecution}. For system calls, \IR{} confirms whether or not they are expected by comparing with the recorded events. For synchronizations, \IR{} confirms whether the address and type of synchronization is expected. Any divergence from the recorded ones can be only caused by unknown race conditions, when all explicit synchronizations and system calls are replayed faithfully. If a divergence is detected, \IR{} will restart a new execution immediately. It will stop performing re-executions when a re-execution with the same events as the recorded ones is found. Currently, \IR{} supports an unlimited number of replays. Note that since \IR{} replays all explicit synchronizations and system calls, it generally takes very few re-executions to find a matching schedule, as evaluated in Table~\ref{table:datarace}. If the replay cannot reproduce the original schedule, \IR{} may insert random delays upon diverging points, but without changing the order of the recorded schedule. Therefore, this mechanism helps reproduce applications with race conditions, as shown in Section~\ref{sec:identicalevaluation}. 

\section{Applications}
\label{sec:applications}

This section presents three example applications built on top of \IR{}: two automatic tools for detecting heap buffer overflows and use-after-free memory errors, and one debugging tool integrating with \texttt{GDB}. The ideas of detecting memory errors are adopted from DoubleTake~\cite{DoubleTake}. These applications exemplify the usefulness of an in-situ and identical record-and-replay system as \IR{}.  

\subsection{Heap Overflow}
\label{sec:application-overflow}

A heap buffer overflow occurs when a program writes outside the boundary of an allocated object.
To aid in error discovery, \IR{} places canaries (e.g., known random values) adjacent to allocated objects in the original execution, a mechanism first introduced by StackGuard~\cite{StackGuard}.
An overflow will corrupt the canary value, which can then be detected at the end of each epoch. Any overwritten canary is incontrovertible evidence that a buffer overflow has occurred. \IR{} uses a bitmap internally to record the placement of canaries.

After the discovery of an overflow, \IR{} immediately triggers a re-execution to locate the exact instructions responsible for the overflow.
Before re-execution, \IR{} installs a watchpoint at every address with a corrupted canary by invoking the \texttt{perf\_event\_open} system call. 
During re-execution, instructions writing to the watched addresses will trigger a trap, such that \IR{} reports the complete call stack of the faulted instruction that causes an overflow. Since there are four watchpoints, \IR{} can identify root causes of four buffer overflows in one re-execution simultaneously. If applications have more than four bugs in one epoch, which is very unlikely in deployed software, \IR{} may invoke multiple replays in order to identify root causes for all bugs.

\subsection{Use-after-free}
\label{sec:use-after-free}

Use-after-free errors occur whenever an application accesses memory that has previously been deallocated, and has possibly been re-allocated to other live objects. A use-after-free error may lead to an immediate \texttt{SIGSEGV} fault, the corruption of data, or other unexpected program behavior. 

To detect use-after-free problems, \IR{} delays the re-allocation of freed objects by placing them into per-thread quarantine lists, an idea originally developed by AddressSanitizer~\cite{AddressSanitizer}. \IR{} fills the first 128 bytes of freed objects with canary values. These freed objects are released from the quarantine list, when the total size of quarantined objects is larger than the user-defined setting. 

\IR{} checks for use-after-free errors before any object is actually removed from the quarantine lists, as well as at epoch boundaries. Similar to buffer overflows, an overwritten canary indicates that a use-after-free error has occurred. \IR{} employs re-executions to identify the root cause of each error, by installing watchpoints at overwritten canaries. During re-execution, \IR{} stores the call stack of allocations and deallocations for the purpose of reporting. It can precisely pinpoint the statements at use-after-free sites by using the watchpoint mechanism. 

\subsection{Interactive Debugging Tool}
\label{sec:debuggingtool}

\IR{} designs an interactive debugging tool to integrate with the \texttt{GDB} debugger in case of abnormal exits, such as assertion failures, segmentation faults, or aborts. \IR{} intercepts these exits and stops inside the signal handler. Therefore, it is possible for programmers to find the call stack associated with abnormal exits, when the process is attached to the debugger. Inside the debugger, programmers may find the addresses of faulted variables, and set watchpoints on these addresses. Afterward, the programmer can issue the rollback command via the debugger, which is supported by \IR{}. For instance, if watchpoints have been set, the \texttt{GDB} debugger will receive notifications when the corresponding addresses have been accessed. Thus, programmers are able to identify the root causes of the fault, without restarting the buggy application. This interactive debugging tool will not only help programmers identify faults in development phases, but can also be utilized in staging deployment~\cite{staging}, especially when new features are rolling out. 

\section{Evaluation}
\label{sec:evaluation}

\subsection{Experimental Setup}
\label{sec:experimentalsetup}

We performed all experiments on a 16-core quiescent machine. This machine has two sockets, installed with Intel(R) Xeon(R) CPU E5-2640 processors and 256GB of memory, and has 256KB L1, 2MB L2, and 20MB L3 cache. The operating system is the vanilla Linux-4.4.25. All applications were compiled using Clang-3.8.1 at the \texttt{-O2} optimization level, except those with explicit explanations.

\paragraph{Evaluated Applications:}
\IR{} was evaluated using 
PARSEC 2.1~\cite{parsec} and several real applications, such as \texttt{mem\-cached 1.4.25}, \texttt{pbzip2}, \texttt{aget}, \texttt{pfscan}, \texttt{Apache {\allowbreak}httpd 2.4.25}, and \texttt{SQlite 3.12.0}.
For PARSEC applications, the native input datasets were used.
\texttt{pbzip2} compressed a 150MB file and \texttt{pfscan} scanned a 826MB file.
\texttt{aget} downloaded 614MB of data from a machine on the same local area network, to avoid interference caused by the Internet.
Memcached was evaluated using a Python script~\cite{memcached}. A program, called ``threadtest3.c'', was used to evaluate {\allowbreak}SQlite~\cite{sqlitetest}.
Apache was evaluated by sending $10,000$ requests via the \texttt{ab} benchmark~\cite{apachetest}. 

\subsection{Identical Re-execution}

\label{sec:identicalevaluation}

We validated the identical execution by checking the order of synchronizations and system calls, as well as the final state of the heap memory. Identical re-executions should always lead to an identical heap image. The probability of a non-identical execution concluding with the same memory state is extremely low. Currently, we did not evaluate explicit outputs, such as file or socket writes, although these evaluations could increase the confidence of the identical execution.  

To perform the validation, we manually implanted a buffer overflow error in the end of \texttt{main} routine for every program. This buffer overflow immediately triggers a re-execution, and we record the memory state before and after the replay. For the default Linux library, we collected the memory differences between these two executions. We also evaluated the memory differences of RR, the only work supporting identical re-execution without special OS support~\cite{RR}.

\begin{table*}[t]
  \centering
  \setlength{\tabcolsep}{0.3em}
  \footnotesize 
    \begin{tabular}{ c|| c | c | c | c | c | c | c | c | c || c |c | c | c | c | c  } 
    \hline
& BS & bodytrack & canneal & dedup & ferret & fluidanimate & streamcluster & swaptions & x264 & aget & apache & memcached & pbzip2 & pfscan & sqlite\\ \hline
\textbf{Orig} & 3 & 9 & 43 & 25 & 5 & 7 & $<0.1$ & 8 & 24 & 7 & 12 & 4 & 4 & 5 & 10\\ \hline
\textbf{IR} & 0 & 0 & 0 & 0 & 0 & 0 & 0 & 0 & 0 & 0 & 0 & 0 & 0 & 0 & 0\\ \hline
\textbf{RR} & 0 & 0 & 0 & 0 & 0 & 0 & 0 & 0 & 0 & 0 & 0 & 0 & 0 & 0 & 0\\

\hline
    \end{tabular}
  \caption{The percentage of memory difference between the original execution and the re-execution \\for the default library (``Orig'') and \IR{} (``IR''). BS is the abbreviation for the blackscholes.  \label{table:memoryoverhead}
  }
    \vspace{-0.1in}
\end{table*}

We evaluated the identical re-execution on 15 applications, including 6 real applications, such as Apache, SQLite, and Memcached. The results of memory differences are listed in Table~\ref{table:memoryoverhead}. Note that \texttt{canneal} cannot be replayed identically initially, since it invokes multiple atomic functions to swap two encapsulated pointers in the original program. As discussed in Section~\ref{sec:intro}, \IR{} cannot support identical replay for applications with ad hoc synchronizations, if without additional instrumentation. Therefore, we manually replaced all atomic instructions (reads/writes) with mutex locks. After these changes, \IR{} achieves the same heap image, as expected. Therefore, both \IR{} and RR can identically reproduce all of these applications, with an identical final heap image at the end. Note that it is much easier for RR to guarantee the identical execution, since it utilizes a single thread to run multiple threads, greatly eliminating the issues caused by concurrency. But, this also indicates that RR cannot be utilized to identify program failures that only occur in concurrent executions.  

\subsubsection{Handling Race Conditions}
In our experiments, \IR{} identically reproduced 14 out of 15 applications (including modified \texttt{canneal}) in their first re-execution, although these applications have more than 146 race conditions in total: \texttt{bodytrack}(10), \texttt{x264}(72), \texttt{streamcluster}(24), \texttt{ferret}(38), and \texttt{pbzip2}(2)~\cite{Effinger-Dean:2012:IIR:2384616.2384650}. 
That is, we did not observe any divergence of the schedule for these 14 applications during their first replay. However, we are not sure regarding whether these races are actually exercised. Only \texttt{bodytrack} requires a second re-execution because of a confirmed race condition related to condition variables~\cite{Effinger-Dean:2012:IIR:2384616.2384650}. Currently, \IR{} does not record the order of condition signal and broadcast, which causes this replay issue. During the second replay, \IR{} successfully reproduces this program by inserting some delays upon diverging points, which avoids the race condition. 

\begin{table}[t!]
  \centering
  \setlength{\tabcolsep}{0.3em}
  \footnotesize 
    \begin{tabular}{c | c |  c | c | c } 
    \hline
    	\specialcell{Replay Times} & 1 & 2 & 3 & $\geq 4$ \\
    \hline
 Percentage & 99.8718 & 0.1088 & 0.0121 & 0.0073 \\ \hline
    \end{tabular}
  \caption{Results of reproducing \texttt{Crasher}'s race. \label{table:datarace}
  }
  \vspace{-0.1in}
\end{table}

In order to further confirm how race conditions affect replay, we also evaluated a synthetic racy program--Crasher~\cite{machado2015concurrency}. We ran Crasher $100,000$ times, and the race condition (causing a crash) was observed on $82,592$ out of $100,000$ executions. Note that normal applications will not have such a high probability of exhibiting a race, since this program intentionally places \texttt{sleep} inside to trigger the race condition. The results of reproducing race conditions is further shown in Table~\ref{table:datarace}. In around $99.87\%$ of executions, \IR{} reproduced the race condition during the first replay. Approximately $0.11\%$ of the time, \IR{} required two replays to reproduce the race, while the remaining ones ($0.02\%$) required more than two replays. These results indicate that \IR{} has an excellent chance to reproduce races, when all other explicit synchronizations are replayed faithfully. 

\subsection{Performance Overhead}
\label{sec:performance}

We compared the performance overhead of \IR{} with \texttt{rr-4.5.0} and CLAP~\cite{Huang:2013:CRL:2491956.2462167}, for the recording phase. \\
\IR{} does not support replay for the whole execution, which is the reason why we did not evaluate performance for the replay phase. 

\texttt{RR} is the only available system supporting identical replay~\cite{RR}.
CLAP is another available RnR system for C/C++ programs that only uses software-based approaches~\cite{Huang:2013:CRL:2491956.2462167}. We cannot find the source code for more recent work, such as Castor~\cite{Castor} and H3~\cite{huang2017towards}. Also, these two recent works actually utilize Intel's Processor Tracing hardware feature, which only appeared after 2013~\cite{pt}. CLAP records thread-local execution paths at runtime, then computes memory dependencies offline. However, their recording mechanism is not available. We re-implemented the recording routine of CLAP based on the path profiling support in LLVM-3.3~\cite{Huang:2013:CRL:2491956.2462167}. Paths were selected by the Ball-Larus algorithm~\cite{Ball:1996:EPP:243846.243857}, and a function call was inserted at the entrance/exit of each function, as well as back edges, in order to record the path number and function call number. Events are recorded in per-thread lists, similar to the design of CLAP.
We have confirmed that the performance of \texttt{aget}, \texttt{pfscan}, and \texttt{bbuf}, with our implemented version, has similar performance to that reported in the paper~\cite{Huang:2013:CRL:2491956.2462167}. We also confirmed the correctness of our implementation with the CLAP authors.   

Results are listed in Table~\ref{table:recoverhead}, where all results are normalized to the runtime of the default \texttt{pthreads} library. \texttt{RR} is compiled with Clang-3.8.1, since it cannot be compiled using the older version of Clang. Applications using CLAP and \IR{} are compiled using Clang-3.3 for fair comparison, since the implementation of the Ball-Larus algorithm is not available after Clang-3.3. CLAP cannot run four applications due to analysis errors in LLVM's path profiling support, and \texttt{RR} cannot run on Apache. 

On average, CLAP runs around $2.4\times$ slower, while \\\IR{} only imposes negligible performance overhead (around 3\%). \texttt{RR} runs around $17\times$ slower, due to using a single thread to run multithreaded programs. In order to investigate why \IR{} behaves better than the default Linux library, we also evaluated the performance of \IR{}'s allocator (noted as ``IR-Alloc'' in Table~\ref{table:recoverhead}). We observed that \IR{}'s custom memory allocator contributes a performance boost of about $2.5\%$, but with worse performance on four evaluated applications. Based on our understanding, there are multiple reasons that can help boost the performance: (1)~the allocator avoids the lock acquisitions of memory allocations/deallocations, since each thread is not sharing the heap with others. (2)~\IR{}'s allocator avoids the large number of \texttt{madvise} system calls (e.g., \texttt{dedup}), and eliminates the possible false sharing effect~\cite{Sheriff}. Thus, the actual recording overhead of \IR{} should be around $6\%$.

For all applications, except \texttt{fluid\-animate} and \texttt{str\-eam\-clus\-ter}, \IR{} introduces less than 10\% re\-cord\-ing overhead. 
Based on our investigation (omitted due to the space limit), \texttt{fluidanimate} has over 54 million lock acquisitions per second, where recording every acquisition and performing the synchronized checking prior to each, adds around 49\% overhead. The overhead of \texttt{streamcluster} mostly comes from \IR{}'s custom memory allocator, for which we do not know the exact reason.  

In contrast, CLAP performs poorly in CPU-intensive applications that have a large amount of back-edges and branches, such as \texttt{ferret}, \texttt{streamcluster}, \texttt{swaptions} and {\allowbreak}\texttt{x264}. For applications that are I/O-intensive (like \texttt{aget}), or applications for which most of their workload is performed in un-instrumented libraries (like \texttt{pbzip2}), CLAP performs very well. \texttt{RR} is typically very slow, except for I/O-bound applications such as \texttt{aget} and \texttt{Memcached}. For other applications, RR is very slow as expected, since it cannot take advantage of the parallelism provided by multiple cores when using only one thread.  

\subsection{Detection Tools}
We also evaluated the effectiveness and performance of detection tools built on top of \IR{}. This evaluation is conducted using applications with real and implanted bugs.  

\subsubsection{Detection Effectiveness}
\label{sec:effectiveness}
We confirmed \IR{}'s effectiveness on heap overflows and use-after-free bugs that were collected from prior tools~\cite{overflow:Cruiser, DoubleTake}, Bugbench~\cite{bugbench}, and Bugzilla~\cite{bzip2overflow}. These applications include \texttt{bc-1.06}, \texttt{bzip2}~\cite{bzip2overflow}, \texttt{gzip-1.2.4}, \texttt{libHX}, \texttt{polymorph}, \texttt{Memcached}~\cite{memcachedbug}, and \\\texttt{libtiff}~\cite{libtiffbug}. \IR{} can detect all of these known problems, which is similar to DoubleTake.  

As described in Section~\ref{sec:identicalevaluation}, we have manually inserted a buffer overflow error at the end of all evaluated applications. \IR{}'s detector can also detect all of these implanted errors. Also, \IR{} reports the root causes of these bugs, with precise calling contexts of the faults.

\subsubsection{Performance Overhead}

\begin{table}[t]
  \centering
  \footnotesize 
    \begin{tabular}{l | r | r | r | r } 
    \hline
      {Applications}
      	& \multicolumn{1}{c|}{IR-Alloc} 
        & \multicolumn{1}{c|}{\IdealReplay} 
        & \multicolumn{1}{c|}{CLAP} 
        & \multicolumn{1}{c}{RR}   
\\
      \hline

blackscholes & 1.001 & 1.021 & 1.113 & 8.011 \\

bodytrack	& 0.993 & 0.990 & - & 27.283 \\

canneal	& 0.880 & 0.962 & - & 6.884 \\

dedup	& 0.664 & 0.817 & 1.074 & 5.138 \\

ferret & 1.017 & 0.998 & 3.519 & 13.275 \\

fluidanimate & 1.044 & 1.493 & 2.183 & 34.082 \\

streamcluster & 1.102 & 1.100 & 2.383 & 52.280 \\

swaptions & 0.979 & 0.990 & 2.964 & 29.921 \\

x264 & 0.991 & 1.032 & 9.100 & 16.228 \\
\hline

aget & 1.000 & 1.032 & 1.013 & 1.065 \\

apache & 1.002 & 1.056 & - & - \\

memcached & 1.000 & 1.001 & 1.001 & 1.822 \\

pbzip2 & 0.974 & 0.895 & - & 26.852 \\

pfscan & 1.015 & 1.006 & 1.032 & 8.462 \\

sqlite & 0.973 & 1.087 & 3.853 & 15.059 \\

\hline
\hline

average & 0.976 & 1.027 & 2.658 & 17.597 \\

\hline
    \end{tabular}
  \caption{Performance overhead\label{table:recoverhead}
  }
  \vspace{-0.1in}
\end{table}

\begin{figure}[!t]
\begin{center}
\includegraphics[width=0.9\columnwidth]{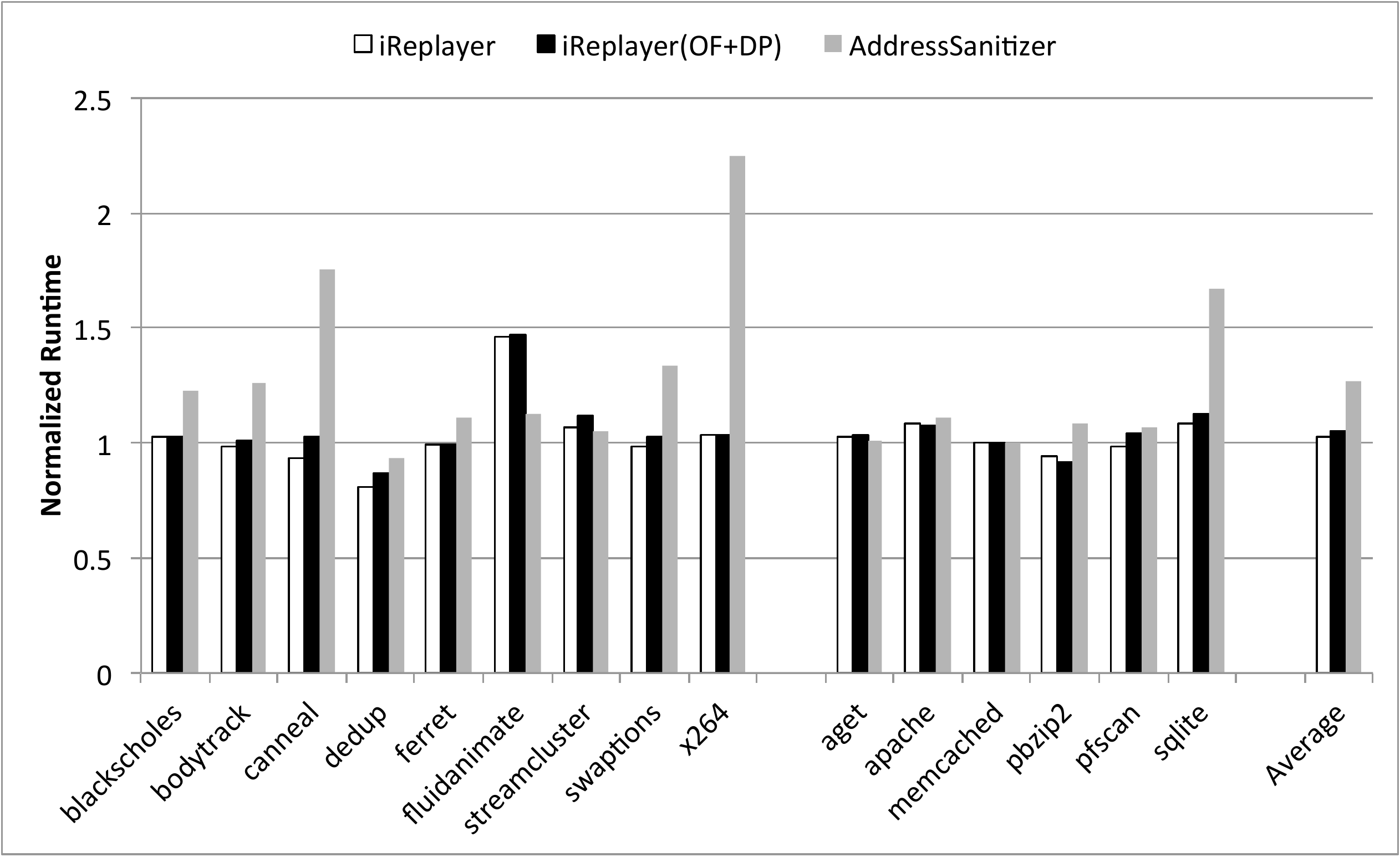}
\end{center}
\vspace{-0.1in}
\caption{
Comparing \IR{}'s performance overhead with AddressSanitizer in detecting memory errors. \label{fig:detectionoverhead}}
\vspace{-0.1in}
\end{figure}

We further compared the performance overhead of \IR{} and its detection tools with AddressSanitizer~\cite{AddressSanitizer}, the previous state-of-the-art in detecting both buffer overflows and use-after-free errors. AddressSanitizer instruments memory accesses during compile time, and checks for possible memory errors by handling instrumented accesses. For a fair comparison, we only enable the instrumentation on memory writes on heap objects, while disabling its leak detection and others. Note that we did not instrument memory writes on all external libraries. We used Clang-3.8.1 for the evaluation, as it ships with the recent version of AddressSanitizer.

As seen in Figure~\ref{fig:detectionoverhead}, \IR{}'s detectors (noted as ``\IR{} (OF+DP)'') only impose around 5\% performance overhead on average, which is significantly lower than that of AddressSanitizer (26\%).
\IR{}'s detectors perform better than, or similar to, AddressSanitizer in almost all applications, except \texttt{fluidanimate}. The overhead of \IR{} on this application comes from the recording of synchronizations, as discussed above. 
It is worth noting that AddressSanitizer cannot detect memory errors caused by non-instrumented components, which includes all external libraries that these applications may invoke. This explains why AddressSanitizer has comparatively much better performance on applications that invoke many non-instrumented libraries, or perform extensive network communications, such as \texttt{aget}, \texttt{Apache}, and \texttt{Mem\-cached}. 

\subsection{Debugging Tools}
We have performed experiments on \texttt{Memcached}, Crash\-er~\cite{machado2015concurrency}, and all evaluated PARSEC applications using implanted buffer overflows. Crasher contains a segmentation fault, while the others have buffer overflows. All of these bugs can be caught using the interactive debugging method, as described in Section~\ref{sec:debuggingtool}.   

\section{Limitations and Future Work}
\label{sec:discussion}

This section discusses some limitations of \IR{}, and possible extensions in the future. 

Firstly, \IR{} only supports epoch-based record-and-{\allowbreak}replay, but not re-execution of the entire program. Re-executing the whole program has better diagnostic capabilities, such as identifying root causes far from the failure site. However, it has some issues as listed in Section~\ref{sec:intro}. There is a chance that replaying the last epoch may miss root cause for some bugs, although existing studies show that most bugs have a very short distance of error propagation and thus should be identifiable~\cite{1209956, Rx, Arulraj:2014:LSM:2541940.2541973}.     

Secondly, it may not achieve identical re-executions when programs with race conditions do not lead to a divergence from the recorded sequence, since it utilizes the order of synchronizations and system calls to determine whether the re-execution is identical to the original one. 
As described above, \IR{} cannot support the identical replay of programs with ad hoc synchronizations, which utilize self-defined synchronizations instead of explicit \texttt{pthreads} APIs. Also, these synchronizations include C/C++ atomics~\cite{atomics}. This issue can be solved by instrumenting the code, as Castor proposed~\cite{Castor}, which allows \IR{}'s runtime to record the order of such events. However, we did not implement this due to two reasons: (1) it requires program instrumentation, which will create barriers for easy deployment. (2) Existing study shows that 22-67\% of ad hoc synchronization uses result in bugs or severe performance issues~\cite{Xiong:2010:AHS:1924943.1924955}, which should be avoided as much as possible.

Thirdly, \IR{}'s detection tools support evidence-based error detection, which cannot detect problems caused exclusively by memory reads, as they do not leave behind evidence of their occurrence. Thus, while they exhibit no false positives, they will miss read-based errors.

\section{Related Work}
\label{sec:relatedwork}

\subsection{Record-and-Replay Systems}
A significant amount of record-and-replay systems exists. We focus on RnR systems that support multithreaded programs with race conditions, and run on the off-the-shelf hardware. 

Some RnR systems require changes to the OS, such as ReVirt~\cite{Dunlap:2002:REI:844128.844148}, Triage~\cite{Tucek:2007:TDP:1294261.1294275}, Respec~\cite{Lee:2010:REO:1736020.1736031}, and DoublePlay~\cite{DoublePlay}, which prevents widespread adoption due to security or reliability concerns related to altering the OS. Some utilize static analysis to reduce runtime overhead, such as ODR~\cite{ODR}, LEAP~\cite{huang2010leap}, CLAP~\cite{Huang:2013:CRL:2491956.2462167}, Light~\cite{Liu:2015:LRV:2737924.2738001}, and H3~\cite{huang2017towards}. However, they may exhibit a scalability issue for their offline analysis. 

Several approaches require no changes to the OS or offline analysis~\cite{Musuvathi:2008:FRH:1855741.1855760, Bhansali:2006:FIT:1134760.1220164, guo2008r2, Patil:2010:PFD:1772954.1772958, Lee:2012:CHP:2254064.2254119, Devecsery:2014:ES:2685048.2685090, Lee:2014:ILR:2945642.2945655, Wang:2014:DDR:2581122.2544152, Castor}, which is closer to \IR{}. However, they also have shortcomings. Some impose more than $10\times$ performance overhead~\cite{Bhansali:2006:FIT:1134760.1220164, Patil:2010:PFD:1772954.1772958}, R2 requires significant manual annotations to specify which functions should be monitored~\cite{guo2008r2}, and some require recompilation to annotate weak-locks on the racy code~\cite{Lee:2012:CHP:2254064.2254119}. 

Similar to \IR{}, some existing work also avoids the recording of racy accesses. Arnold detects the divergence of executions caused by race conditions, and can attach a vector-lock data race detector in replay~\cite{Devecsery:2014:ES:2685048.2685090}. However, it relies on manual instrumentation to fix them. Lee et al. employ multiple tries (based on a single-threaded re-execution) to search for a matching schedule~\cite{Lee:2014:ILR:2945642.2945655}, while \IR{} employs multiple threads to replay, and can find a matched schedule in fewer tries. Castor utilizes the hardware synchronized timestamp counters to order events, and hardware transactional memory to reduce locking overhead inside critical sections~\cite{Castor}. Thus, Castor requires hardware support and compiler instrumentation that prevents its easy deployment.
\IR{} does not rely on any hardware feature, but employs a novel data structure and level of indirection to avoid significant recording overhead on synchronizations. Overall, \IR{} achieves a similar level of performance overhead as Castor, but can identically reproduce programs without ad hoc synchronizations. 

\paragraph{Difference between \IR{} and RR:} No existing {\allowbreak}work can guarantee identical re-execution in the in-situ setting. \texttt{RR} is the only available system that supports identical re-execution in the offline setting~\cite{RR}. However, \texttt{RR} executes and replays multiple threads using a time-sharing method on a single core, which makes it easier to achieve identical replay. Due to the lack of scalability to multicore hardware, \texttt{RR} runs more than $17\times$ slower. Further, \texttt{RR} does not support in-situ replay. By comparison, \IR{}'s overhead is less than $3\%$, and supports the in-situ replay. 

\subsection{Deterministic Multithreading} 
Deterministic multithreading (DMT) systems is another interesting direction that is distinct from this work~\cite{Bergan:2010:CCR:1736020.1736029, Aviram:2010:ESD:1924943.1924957, Dthreads, Cui:2011:EDM:2043556.2043588, Cui:2013:PPR:2517349.2522735, Devietti:2011:RRC:1950365.1950376}. DMT systems are generally unsuitable for debugging purposes, as they can only exercise one possible schedule. Although they completely avoid recording overhead by always enforcing a deterministic order on synchronizations, they may impose much larger performance overhead when handling race conditions~\cite{Cui:2013:PPR:2517349.2522735}.  

\subsection{Detecting Memory Errors} 
Many dynamic approaches can detect memory errors, since they do not generate false positives. Typically, they utilize either dynamic or static instrumentation to instrument every memory access. Dynamic instrumentation tools do not require the recompilation of source code~\cite{overflow:drmemory, overflow:purify, overflow:inspector, overflow:valgrind, overflow:discover}, but may increase performance overhead as high as an order of magnitude. Static instrumentation tools may employ static analysis to help reduce the volume of instrumentation~\cite{overflow:Baggy, overflow:Mudflap, overflow:lbc, overflow:ccured, overflow:Insure++, AddressSanitizer}. AddressSanitizer is the state-of-the-art in dynamic analysis tools~\cite{AddressSanitizer}, which can also detect out-of-bound reads on stack and global variables that is available with \IR{}. However, AddressSanitizer requires explicit instrumentation, and cannot be utilized in a production environment due to its prohibitive performance overhead. DoubleTake provides similar functionality to \IR{}~\cite{DoubleTake}. However, it cannot support multithreaded programs, which is very challenging to achieve, as discussed in Section~\ref{sec:overview}. Also, DoubleTake does not support the same interactive debugging as that of \IR{}.

\section{Conclusion}
\label{sec:conclusion}

This paper introduced \IR{}, a novel system that supports identical replay in the in-situ setting. \IR{} imposes only approximately $3\%$ recording overhead, and can identically reproduce all applications without ad hoc synchronizations. To demonstrate its usefulness, three tools are built on top of it: two automatic tools for detecting the root causes of buffer overflows and use-after-free errors, and an interactive debugging tool that helps identify the source of segmentation faults and other abnormal exits. 

\section*{Acknowledgment}

This work was initiated and partially conducted while Tongping Liu was a Ph.D. student at the University of Massachu\-setts Amherst, under the supervision of Professor Emery Berger. We would like to thank our shepherd, Yannis Smaragdakis, and anonymous reviewers for their valuable suggestions and feedback. We appreciate Charlie Curtsinger, Bobby Powers, and Jinpeng Zhou for their participation in development, and Xiangyu Zhang, Michael D. Bond, Jia Rao, Corey Crosser, and Mary Mays for their helpful comments. We also thank Jeff Huang, Kostya Serebryany, Nuno Machado, Jason Flinn, Brandon Lucia, Kaushik Veeraraghavan, and David Devecsery for their help on the evaluation. This material is based upon work supported by the National Science Foundation under Award CCF-1566154 and CCF-1617390. The work is also supported by Google Faculty Award, Mozilla Research Grant, and the startup package from University of Texas at San Antonio. 


{
\bibliographystyle{acm}
\bibliography{ref,refs,reviewer,tongping}
}

\end{document}